\documentclass[aps,prd,twocolumn,superscriptaddress,nofootinbib]{revtex4-1}

\usepackage{latexsym}
\usepackage{amsmath}
\usepackage{amssymb}
\usepackage{amsfonts}
\usepackage{bm}
\usepackage{physics}
\usepackage[table,x11names]{xcolor}
\usepackage{color}
\definecolor{purple}{rgb}{0.5,0,0.5}
\definecolor{blue}{rgb}{0.0,0,0.9}
\definecolor{prdblue}{rgb}{0.133,0.118,0.498}
\usepackage[colorlinks=true, pdfstartview=FitV, linkcolor=prdblue, citecolor= prdblue, urlcolor=prdblue]{hyperref}

\usepackage{supertabular} 
\usepackage{placeins}
\usepackage{epsfig}
\usepackage{graphicx}

\begin{document}


\title{Exploring $T_{\psi\psi}$ tetraquark candidates in a coupled-channels formalism}


\author{P. G. Ortega}
\email[]{pgortega@usal.es}
\affiliation{Departamento de F\'isica Fundamental, Universidad de Salamanca, E-37008 Salamanca, Spain}
\affiliation{Instituto Universitario de F\'isica 
Fundamental y Matem\'aticas (IUFFyM), Universidad de Salamanca, E-37008 Salamanca, Spain}

\author{D. R. Entem}
\email[]{entem@usal.es}
\affiliation{Departamento de F\'isica Fundamental, Universidad de Salamanca, E-37008 Salamanca, Spain}
\affiliation{Instituto Universitario de F\'isica
Fundamental y Matem\'aticas (IUFFyM), Universidad de Salamanca, E-37008 Salamanca, Spain}

\author{F. Fern\'andez}
\email[]{fdz@usal.es}
\affiliation{Instituto Universitario de F\'isica 
Fundamental y Matem\'aticas (IUFFyM), Universidad de Salamanca, E-37008 Salamanca, Spain}

\date{\today}

\begin{abstract}
This study investigates the properties of the $T_{\psi\psi}$ tetraquark candidates within a coupled-channels calculation of the $c\bar c- c\bar c$ system, specifically focusing on the $J^P=0^\pm$, $1^\pm$, and $2^\pm$ sectors. The analysis includes various channels containing a $J/\psi$, $\psi^\prime$, $\eta_c$, and $\eta_c^\prime$ meson. By searching for poles in the scattering matrix, a total of 29 states in different $J^P$ sectors with masses ranging from 6.1 to 7.6 GeV/c$^2$ are identified. The study further investigates the masses, widths and branching ratios of these states, leading to the identification of two potential candidates for the experimental $T_{\psi\psi}(6200)$ tetraquark, one candidate for $T_{\psi\psi}(6600)$, two for $T_{\psi\psi}(6700)$, four for $T_{\psi\psi}(6900)$, and three for $T_{\psi\psi}(7200)$ tetraquarks. Additionally, the paper discusses strategies to discriminate between different candidates and explores possible detection channels for further $c\bar c- c\bar c$ states.
\end{abstract}


\maketitle


\section{INTRODUCTION}
\label{sec:introduction}

Understanding the spectroscopy, structure and dynamics of exotic hadrons is one of the most challenging areas of contemporary physics research. In recent years, high-energy experiments have revealed a wealth of multiquark states that defy conventional explanations based on baryon ($qqq$) or meson ($q\bar q$) configurations. The seminal discovery of X(3872) by the Belle group~\cite{Belle:2003nnu} marked a turning point, subsequently leading to the identification of several tetraquark states, including $Z_c(3900)$, $Z_c(4020)$~\cite{BESIII:2013ris,BESIII:2013ouc}, $Z_{cs}(3985)^-$, $Z_{cs}(4220)^+$~\cite{BESIII:2020qkh,LHCb:2021uow}, which exhibit charmonium-like properties, $Z_b(10610)$, $Z_b(10650)$~\cite{Belle:2011aa,Belle:2015upu}, which resemble bottomonium states, or openly exotic states such as the $T_{cc}(3875)^+$~\cite{LHCb:2021vvq,LHCb:2021auc} or the $T_{cs0}(2900)^0$, $T_{cs1}(2900)^0$~\cite{LHCb:2020pxc} particles. The study of the properties and behaviour of these exotic hadrons promises to deepen our understanding of the fundamental interactions that govern the subatomic world, transcending the conventional quark compositions.

Recent breakthroughs have been made by the LHCb, CMS, and ATLAS Collaborations, as they have observed resonances in the di-$J/\psi$ and $J/\psi \psi^\prime$~\footnote{For simplicity, in this work we will denote $\psi(2S)$ as $\psi^\prime$ and $\eta_c(2S)$ as $\eta_c^\prime$.} invariant mass distributions~\cite{LHCb:2020bwg,CMS:2023owd,Zhang:2022toq,ATLAS:2022hhx,Xu:2022rnl} in proton-proton collision data at $\sqrt{s}=7$, $8$ and $13$ TeV. These resonances, with $cc\bar c\bar c$ minimum quark content, such as $T_{\psi\psi}(6200)$, $T_{\psi\psi}(6600)$, $T_{\psi\psi}(6900)$ and $T_{\psi\psi}(7200)$~\footnote{In this work we will follow the naming convention of Ref.~\cite{Gershon:2022xnn}.} has sparked renewed interest in investigating fully charmed and beauty four-quark mesons. These experimental results provide a unique opportunity to test and refine our current understanding in this field.

The existence of heavy exotic mesons composed of two or four $c$ and $b$ quarks has intrigued researchers since the early stages of multiquark hadron studies~\cite{Iwasaki:1976cn,Chao:1980dv,Ader:1981db,Badalian:1985es,Lloyd:2003yc,Berezhnoy:2011xn,Karliner:2016zzc} and, since the experimental observation of $T_{\psi\psi}$ candidates a large number of theoretical studies have been devoted to explaining their properties, either as compact tetraquark states~\cite{Park:2018wjk,Di:2018dcf,Lu:2020cns,Karliner:2020dta,Weng:2020jao,Sonnenschein:2020nwn,Gordillo:2020sgc}, diquark-antidiquark structures~\cite{Debastiani:2017msn,Chen:2016jxd,Wang:2019rdo,Bedolla:2019zwg,Giron:2020wpx,Jin:2020jfc,Deng:2020iqw,Faustov:2021hjs,Wang:2020ols,Mutuk:2021hmi} and meson-meson molecules or coupled-channels effects~\cite{Debastiani:2017msn,Dong:2020nwy,Jin:2020jfc,Deng:2020iqw,Albuquerque:2020hio,Guo:2020pvt,Agaev:2023ruu,Niu:2022jqp}.

Many of these exotic states, such as $X(3872)$, $Z_b(10610)$, $Z_c(3900)$, $P_c(4470)$ and others, tend to emerge close a two-hadron threshold. It is therefore tempting to infer a molecular nature for such kind of states. Similarly, many of the recent $T_{\psi\psi}$ states such as the $T_{\psi\psi}(6200)$, $T_{\psi\psi}(6600)$ or the $T_{\psi\psi}(6900)$, are close to many charmonium-charmonium thresholds such as the $J/\psi J/\psi$, $\eta_c\eta_c^\prime$ or the $J/\psi \psi^\prime$ threshold, respectively. Motivated by these observations, this study investigates the properties of the $T_{\psi\psi}$ candidates $T_{\psi\psi}(6200)$, $T_{\psi\psi}(6600)$, $T_{\psi\psi}(6700)$, $T_{\psi\psi}(6900)$ and $T_{\psi\psi}(7200)$ in a coupled-channels formalism based on a constituent quark model (CQM)~\cite{Vijande:2004he, Segovia:2013wma}, which has been widely used in the heavy quark sector~\cite{Segovia:2008zz,Segovia:2016xqb, Ortega:2020uvc} and extended to the study of other exotic states such as the $X(3872)$~\cite{Ortega:2009hj, Ortega:2016pgg, Ortega:2018cnm, Ortega:2021xst, Ortega:2021fem}, the $T_{cc}^+$~\cite{Ortega:2022efc} or the $T_{cs}$ and $T_{c\bar s}$ states~\cite{Ortega:2023azl}. The advantage of using an approach with a relatively long history is that all model parameters are already constrained by previous works. Consequently, from this point of view, we present a parameter-free calculation of the $T_{\psi\psi}$ states, extending our recent analysis of the similar $T_{cc}^+$ and $T_{cs}$ exotic candidates~\cite{Ortega:2022efc,Ortega:2023azl}.

The organization of the manuscript is as follows: After this introduction, section~\ref{sec:theory} provides a brief overview of the theoretical framework. Section~\ref{sec:results} primarily focuses on the analysis and discussion of our theoretical findings. Lastly, in Sec.~\ref{sec:summary}, we present a summary of our work and draw conclusions based on the obtained results.


\section{THEORETICAL FORMALISM}
\label{sec:theory} 

In this work we will explore the $T_{\psi\psi}$ tetraquark candidates as meson-meson molecules. This system has many similar features as  the recently discovered $T_{cc}^+$ tetraquark, with minimum quark content $cc\bar u\bar d$. Then, for the $T_{\psi\psi}$ we will follow the same formalism as in Ref.~\cite{Ortega:2022efc}, where the $T_{cc}^+$ was described as a $J^P=1^+$ $DD^*$ molecule. For this reason, in this section we will only briefly provide the most relevant theoretical aspects for the study of the $T_{\psi\psi}$ states.

The constituent quark model (CQM) employed in this work has been extensively detailed in the literature. For a full description, including expressions of all the potentials and the values of the model parameters, the reader is kindly referred to Ref.~\cite{Vijande:2004he} and its update Ref.~\cite{Segovia:2008zz}. 

The main elements of our constituent quark model (CQM) encompass the constituent light quark masses and the exchanges involving Goldstone bosons, which arise as manifestations of the dynamical breaking of chiral symmetry in Quantum Chromodynamics (QCD). Additionally, the model incorporates the perturbative interaction of one-gluon exchange (OGE) and a non-perturbative confinement interaction~\cite{Vijande:2004he, Segovia:2013wma}.
 However, it is worth noticing that, whereas the Goldstone boson exchanges are considered for two light quarks ($qq$), they are not allowed in the light-heavy ($qQ$) and heavy-heavy ($QQ$) configurations.\footnote{Here, we denote $q=\{u,d,s\}$ and $Q=\{c,b\}$.} On the contrary, the most important contributions of the one-gluon exchange and confinement potentials are flavour-blind and are the only interactions relevant for this work, where all the quarks involved are beyond the chiral symmetry breaking scale.
 
Regarding the confinement interaction, while it has been proven that multi-gluon exchanges generate an attractive potential that rises linearly with the distance between infinitely heavy quarks~\cite{Bali:2000gf}, it is essential to consider the influence of sea quarks on the strong interaction dynamics. Sea quarks contribute to screening the rising potential at low momenta and eventually lead to the breaking of the quark-antiquark binding string~\cite{Bali:2005fu}. To account for this behaviour, our CQM incorporates the following expression:

\begin{equation}
V_{\rm CON}(\vec{r}\,)=\left[-a_{c}(1-e^{-\mu_{c}r})+\Delta \right] (\vec{\lambda}_{q}^{c}\cdot\vec{\lambda}_{\bar{q}}^{c}) \,,
\label{eq:conf}
\end{equation}
where $a_{c}$ and $\mu_{c}$ are model parameters. At short distances this potential exhibits a linear behavior with an effective confinement strength, $\sigma=-a_{c}\,\mu_{c}\,(\vec{\lambda}^{c}_{i}\cdot \vec{\lambda}^{c}_{j})$. However, it becomes constant at large distances, with a threshold defined by $\{\Delta-a_{c}\}(\vec{\lambda}^{c}_{i}\cdot \vec{\lambda}^{c}_{j})$.

Additionally, the model incorporates QCD perturbative effects mediated by the exchange of one gluon, derived from the vertex Lagrangian
\begin{equation}
{\mathcal L}_{qqg} = i\sqrt{4\pi\alpha_{s}} \, \bar{\psi} \gamma_{\mu} 
G^{\mu}_c \lambda^c \psi \,.
\label{eq:Lqqg}
\end{equation}
Here, $\alpha_{s}$  represents an effective scale-dependent strong coupling constant, given by

\begin{equation}
 \alpha_s(\mu)=\frac{\alpha_0}{\ln\left(\frac{\mu^2+\mu_0^2}{\Lambda_0^2}\right)}
\end{equation}
where $\mu$ is the reduced mass of the $q\bar q$ pair and $\alpha_0$, $\mu_0$ and $\Lambda_0$ are parameters of the model~\cite{Segovia:2013wma}.

The described CQM details the $qq$ ($q\bar q$) interaction at microscopic level and allows us to build the $c\bar c$ meson spectra~\cite{Segovia:2013wma,Segovia:2008zz}, by solving the two-body Schr\"odinger equation through the use of the Gaussian Expansion Method~\cite{Hiyama:2003cu}. This computational approach not only simplifies the evaluation of the necessary matrix elements but also ensures a satisfactory level of accuracy. 

In order to describe the $c\bar c-c\bar c$ interaction from the underlying $qq$ dynamics we employ the Resonating Group Method~\cite{Wheeler:1937zza}. 
For that, we assume that the wave function of a system composed of two charmonium mesons $A$ and $B$ can be written as

\begin{align}\label{eq:wf}
 \Psi = {\cal A}\left[\phi_{A}\phi_{B}\chi_L\sigma_{ST}\xi_c\right]
\end{align}
where $\phi_{A(B)}$ is the wave functions of the $A(B)$ meson, $\chi_L$ the relative orbital wave function of the $AB$ pair, $\sigma_{ST}$ their spin-isospin wave function and $\xi_c$ their color wave function.

As we have two pair of identical quarks, we have to consider the full antisymmetric operator ${\cal A}$, so the wave function is completely antisymmetric.
For the $c\bar c-c\bar c$ system, this operator can be written as ${\cal A}=(1-P_c)(1-P_{\bar c})$, up to a normalization factor, where $P_c$ is the operator that exchanges $c$ quarks and $P_{\bar c}$ the operator that exchanges charm antiquarks between mesons. Following Ref.~\cite{Ortega:2022efc}, for identical mesons, the antisymmetrizer is reduced to $\Psi = (1-P_{\bar c}) \left\{|\phi_{A} \phi_{B}\chi_{L}\sigma_{ST}\xi_c \rangle\right\}$, whereas for non-identical mesons, the wave functions is a combination of $AB$ and $BA$ configurations, given by
\begin{align}\label{eq:antisym}
\Psi &= (1-P_{\bar c}) \left\{
|\phi_{A} \phi_{B}\chi_{L} \sigma_{ST}\xi_c\rangle
+(-1)^\mu |\phi_{B} \phi_{A}\chi_{L}\sigma_{ST}\xi_c \rangle \right\}
\end{align}
with $\mu=L+S-J_A-J_B$. 

As the charmonium states are eigenstates of the $C$-parity operator, the $C$ parity of the $AB$ pair is defined as $(-1)^{L_A+S_A+L_B+S_B}$.
Hence, it is equal to $C=1$ for $PP$ and $VV$ channels (where $P$ is a pseudoscalar meson and $V$ a vector meson) and $C=-1$ for $PV$ channels.

The interaction between $c\bar c-c\bar c$ mesons can be split into a direct term, with no quark exchange between clusters, and an exchange kernel, which incorporates them. The direct potential $V_{D}(\vec{P}',\vec{P}_{i})$ can be written as
\begin{align}
&
V_{D}(\vec{P}',\vec{P}_{i}) = \sum_{i\in A, j\in B} \int d\vec{p}_{A'} d\vec{p}_{B'} d\vec{p}_{A} d\vec{p}_{B} \times \nonumber \\
&
\times \phi_{A'}^{\ast}(\vec{p}_{A'}) \phi_{B'}^{\ast}(\vec{p}_{B'})
V_{ij}(\vec{P}',\vec{P}_{i}) \phi_{A}(\vec{p}_{A}) \phi_{B}(\vec{p}_{B})  \,,
\end{align}
where $V_{ij}$ is the CQM potential between the quark $i$ and the quark $j$ of the mesons $A$ and $B$, respectively.

 The exchange kernel $K_E$, that models the quark rearrangement between clusters, can be written as
\begin{align}
K_E(\vec P',\vec P_i) &= H_E(\vec P',\vec P_i) - E_T\, N_E(\vec P',\vec P_i)\,.
\end{align}
which is a non-local and energy-dependent kernel, separated into a potential term $H_E$ plus a normalization term $N_E$. Here, $E_T$ denotes the total energy of the system and $\vec P_i$ is a continuous parameter. The exchange Hamiltonian and normalization can be written as
\begin{subequations}
\begin{align}\label{eq:exchangeV}
&
H_{E}(\vec{P}',\vec{P}_{i}) = \int d\vec{p}_{A'}
d\vec{p}_{B'} d\vec{p}_{A} d\vec{p}_{B} d\vec{P} \phi_{A'}^{\ast}(\vec{p}_{A'}) \times \nonumber \\
&
\times  \phi_{B'}^{\ast}(\vec{p}_{B'})
{\cal H}(\vec{P}',\vec{P}) P_{\bar c} \left[\phi_A(\vec{p}_{A}) \phi_B(\vec{p}_{B}) \delta^{(3)}(\vec{P}-\vec{P}_{i}) \right] \,,\\
&
N_{E}(\vec{P}',\vec{P}_{i}) = \int d\vec{p}_{A'}
d\vec{p}_{B'} d\vec{p}_{A} d\vec{p}_{B} d\vec{P} \phi_{A'}^{\ast}(\vec{p}_{A'}) \times \nonumber \\
&
\times  \phi_{B'}^{\ast}(\vec{p}_{B'})
P_{\bar c} \left[\phi_A(\vec{p}_{A}) \phi_B(\vec{p}_{B}) \delta^{(3)}(\vec{P}-\vec{P}_{i}) \right] \,,
\end{align}
\end{subequations}
where ${\cal H}$ is the Hamiltonian at quark level.

The properties of the $T_{\psi\psi}$ tetraquark candidates, investigated here as meson-meson molecular systems, will be obtained as poles of the scattering matrix, given in non-relativistic kinematics as,

\begin{equation}
S_{\alpha}^{\alpha'} = 1 - 2\pi i
\sqrt{\mu_{\alpha}\mu_{\alpha'}k_{\alpha}k_{\alpha'}} ,
T_{\alpha}^{\alpha'}(E+i0^{+};k_{\alpha'},k_{\alpha}) ,
\end{equation}
where $k_{\alpha}$ and $\mu_\alpha$ represents the on-shell momentum and reduced mass for channel $\alpha$, respectively.
The $T$ matrix of the coupled-channels calculation is obtained  from the Lippmann-Schwinger equation
\begin{align} \label{ec:Tonshell}
T_\beta^{\beta'}(z;p',p) &= V_\beta^{\beta'}(p',p)+\sum_{\beta''}\int dq\,q^2\,
V_{\beta''}^{\beta'}(p',q) \nonumber \\
&
\times \frac{1}{z-E_{\beta''}(q)}T_{\beta}^{\beta''}(z;q,p) \,,
\end{align}
where $\beta$ represents the set of quantum numbers necessary to determine a partial wave in the meson-meson channel, $V_{\beta}^{\beta'}(p',p)$ is the full RGM potential, sum of direct and exchange kernels, and  $E_{\beta''}(q)$ is the energy for the momentum $q$ referred to the lower threshold.


\begin{table*}[t!]
\caption{\label{tab:thres} Channels evaluated in the coupled-channels calculation of this work, along with the included partial waves of each channel in different $J^P$ sectors, denoted as $^{2S+1}L_J$.}
\begin{ruledtabular}
\begin{tabular}{cccccccc}
 Channel & Mass & $0^-$ & $0^+$ & $1^-$ & $1^+$ & $2^-$ & $2^+$  \\\hline
$\eta_c\eta_c$ &    $5966.8$ & - & $^1S_0$ & - & - & - & $^1D_2$ \\
$\eta_cJ/\psi$ &    $6080.3$ & $^3P_0$ & - & $^3P_1$ & $^3S_1$ - $^3D_1$ & $^3P_2$ & $^3D_2$\\
$J/\psi J/\psi$ &    $6193.8$ & $^3P_0$ & $^1S_0$ - $^5D_0$ & $^3P_1$ & - & $^3P_2$ & $^5S_2$ - $^1D_2$ - $^5D_2$ \\
$\eta_c\eta_c^\prime $ &   $6622.6$ & - & $^1S_0$ & $^1P_1$ & - & - & $^1D_2$\\
$\eta_c\psi^\prime $ &    $6669.5$ & $^3P_0$ & - & $^3P_1$ & $^3S_1$ - $^3D_1$ & $^3P_2$ & $^3D_2$\\
$\eta_c^\prime J/\psi$ &   $6736.1$ & $^3P_0$ & - & $^3P_1$ & $^3S_1$ - $^3D_1$ & $^3P_2$ & $^3D_2$\\
$J/\psi\psi^\prime $ &    $6783.0$& $^3P_0$ & $^1S_0$ - $^5D_0$ & $^1P_1$ - $^3P_1$ - $^5P_1$ & $^3S_1$ - $^3D_1$ & $^3P_2$ & $^5S_2$ - $^1D_2$ - $^3D_2$ - $^5D_2$ \\
$\eta_c^\prime \eta_c^\prime $  &  $7278.4$ & - & $^1S_0$ & - & - & - & $^1D_2$\\
$\eta_c^\prime \psi^\prime $  &    $7325.3$ & $^3P_0$ & - & $^3P_1$ & $^3S_1$ - $^3D_1$ & $^3P_2$ & $^3D_2$\\
$\psi^\prime \psi^\prime $ &    $7372.2$ & $^3P_0$ & $^1S_0$ - $^5D_0$ & $^3P_1$ & - & $^3P_2$ & $^5S_2$ - $^1D_2$ - $^5D_2$ \\
\end{tabular}
\end{ruledtabular}
\end{table*}

The mass and the total width of resonances can be directly obtained from the complex energy of the poles, $\bar E=M_r-i\,\frac{\Gamma_r}{2}$.
However, some caution should be taken in order to obtain the partial widths of the resonances to a specific final meson-meson channel. For that, we will follow Refs.~\cite{Ortega:2012rs,Grassi:2000dz}. In the neighborhood of a resonance, the $S$ matrix can be approximated as
\begin{align}
S^{\beta'\beta}(E)=&
S^{\beta'\beta}_{bg}(E)
- i2\pi\delta^{4}(P_f-P_i)
\dfrac{g^{\beta'}g^{\beta}}{E-\bar E}
\end{align}
where $g^{\beta}$ are the residues of the pole, which can be interpreted as the amplitude of the resonance to the final state. The partial width of the resonance to the final state $f$ can be defined as
\begin{equation} 
\label{ec:parcial1}
\hat \Gamma_f=\int d\Phi_f |S(X\rightarrow f)|^2
\end{equation}
where the integral is over the phase space of the final state with 
$\left(\sum_n p_n\right)^2=M_r^2$, with $M_r$ the mass of the resonance.
In the case of a two meson decay, $\hat \Gamma_\beta$ can be written as
\begin{align}\label{ec:parcial2}
 \hat\Gamma_\beta=&2\pi \dfrac{E_1E_2}{M_r}{k_0}_\beta  |g^{\beta}|^2
\end{align}
where ${k_0}_\beta$ is the relativistic onshell momentum of the final two meson state.

It is worth noticing that Eq.~\eqref{ec:parcial2} does not guarantee that the sum of the partial 
widths must be equal to the total width. In fact, it is expected that $\sum_f
\hat \Gamma_f\neq \Gamma_r$. To solve this problem we define the
$\emph{branching ratios}$ as~\cite{Grassi:2000dz}
\begin{equation} \label{ec:bratios}
\mathcal{B}_f=\frac{\hat\Gamma_f}{\sum_{f'}\hat \Gamma_{f'}´}
\end{equation}
so the physical partial widths are given, as usual, by
\begin{equation} 
  \Gamma_f=\mathcal{B}_f \Gamma_r.
\end{equation}
with $\Gamma_r=-2\Im(\bar E)$.

\section{RESULTS}
\label{sec:results}

In this section we present the results of the coupled-channels calculation of the $c\bar c-c\bar c$ system in $J^P=0^\pm,1^\pm,2^\pm$. 
We have included the channels and partial waves shown in Table~\ref{tab:thres}, which are the combination of the lowest lying $S$-wave charmonium resonances, that's it: $J/\psi$, $\psi^\prime $, $\eta_c$ and $\eta_c^\prime $. We restrict ourselves to relative orbital momenta $L\le 2$, since higher ones are expected to be negligible.

Direct interactions are only driven by gluon annihilation diagrams, which are rather small for charmonium. Confinement potential does not have direct interaction because we deal with a two-color-singlet system. Thus, the leading interaction is the exchange diagrams. This implies that their identification as pure molecules is questionable as we are not dealing with a residual direct interaction, but a short-range interaction that mixes quarks. Nevertheless, in this work we will denote the found states as molecules, in a broad sense of a resonant state of two colourless mesons, regardless of the binding mechanism. 

Before presenting the results, it is worth mentioning that there is a theoretical uncertainty in the results as a consequence on the way the model parameters are adjusted to describe a certain number of hadron observables. Such fitting is done within a determinate range of agreement with the experiment, which is estimated to be around 10-20\% for physical observables that help to fix the model parameters. This range of agreement will be taken as an estimate of the model uncertainty for the derived quantities and, in order to analyse its effect, we will estimate the error of the pole properties by varying the strength of the potentials by $\pm10\%$.

The results of our calculations are shown in Table~\ref{tab:Tjj2} (masses, widths and branching ratios). We find up to 29 poles in different $J^P$ sectors, that's it: $2$ in $0^-$, $9$ in $0^+$, $5$ in $1^-$, $5$ in $1^+$, $2$ in $2^-$ and $6$ in $2^+$. Their masses range from $6.1$ to $7.6$ GeV and are quite broad. Due to Heavy Quark Spin Symmetry, the states are relatively degenerate between the $\{0^-,1^-,2^-\}$ and the $\{0^+,1^+,2^+\}$ sectors, but there are significant deviations due to the specific partial waves on each sector.

The most explored detection channels are $J/\psi J/\psi$ and $J/\psi \psi^\prime$. In Table~\ref{tab:Tjj2} we can identify up to $13$ states with significant branching ratios to the $J/\psi J/\psi$ channel, and another $12$ states that can decay to the $J/\psi \psi^\prime$ channel. Among them, we can identify candidates for the experimental states $T_{\psi\psi}(6200)$, $T_{\psi\psi}(6600)$, $T_{\psi\psi}(6700)$, $T_{\psi\psi}(6900)$ and $T_{\psi\psi}(7200)$, which are described in more detail below.

Additionally, we have candidates that do not decay to the above channels. For example, the two $0^{--}$ and $2^{--}$ wide resonances with masses around $6740$ MeV/c$^2$ decay only to $\eta_c \psi^\prime$, while the two $0^{++}$ and $1^{+-}$ states with masses around $6100$ MeV/c$^2$ can only decay to $\eta_c\eta_c$ and $\eta_c J/\psi$, respectively. We also find a broad resonance in the $1^{--}$ sector with a mass of $6822_{-4}^{+3}$ MeV/c$^2$ and a width of $405_{-16}^{+18}$ MeV, which decays mostly to $\eta_c\psi^\prime$ and $\eta_c^\prime J/\psi$. Recently, Belle Collaboration searched for double-charmonium states in the $e^+e^-\to\eta_cJ/\psi$ reaction and found no significant signal~\cite{Belle:2023gln}. This is consistent with our results and points to $e^+e^-\to \eta_c\psi^\prime$ and $e^+e^-\to \eta_c^\prime J/\psi$ as more promising reactions.

\subsection{$T_{\psi\psi}(6200)$}

The $T_{\psi\psi}(6200)$ (or $T_{\psi\psi}(6220)$) tetraquark was discovered in ATLAS~\cite{ATLAS:2022hhx} in the $J/\psi J/\psi$ channel, but its existence was previously suggested in Ref.~\cite{Dong:2020nwy} from an \-a\-na\-ly\-sis of the near-threshold region of the $J/\psi J/\psi$ invariant mass spectrum measured by LHCb~\cite{LHCb:2020bwg}. Its mass and width is $6220\pm50$ MeV/c$^2$ and $310\pm120$ MeV, respectively. Its quantum numbers are not yet determined, but Ref.~\cite{Dong:2020nwy} argued it as a $0^{++}$ or $2^{++}$ $J/\psi J/\psi$ structure.
Other theoretical studies give similar predictions. For example, Ref.~\cite{Agaev:2023ruu} assign the $T_{\psi\psi}(6200)$ state as a $\eta_c\eta_c$ $0^{++}$ molecule using the QCD sum rule method, 
Ref.~\cite{Wang:2022xja} supported its assignment as a ground state tetraquark with $J^{PC}=0^{++}$ or $1^{+-}$, 
Ref.~\cite{Dong:2022sef} identifies it as the $0^{++}$ tetraquark, same as Ref.~\cite{Faustov:2022mvs} though the authors also have a near $1^{+-}$ candidates.
Ref.~\cite{Weng:2020jao} predicts tetraquark states close to $6.2$ GeV/c$^2$ at $0^{++}$, $1^{+-}$ and $2^{++}$, Ref.~\cite{Chen:2016jxd} have close candidates in $1^{++}$, $1^{+-}$ and $2^{++}$, Ref.~\cite{Bedolla:2019zwg} in $1^{+-}$ and $2^{++}$, Ref.~\cite{Jin:2020jfc} in $0^+$ and $2^+$ and Ref.~\cite{Albuquerque:2020hio} describe them as a $0^{++}$ tetraquark state.

In our coupled-channels calculation we find three possible candidates near the mass of the experimental $T_{\psi\psi}(6200)$ in the $J^{PC}=0^{++}$, $1^{+-}$ and $2^{++}$ sectors. However, in the $1^{+-}$ sector the detection channel $J/\psi J/\psi$ is closed and it only decays to the $\eta_c J/\psi$ channel, so only the $0^{++}$ and $2^{++}$ candidates can decay to $J/\psi J/\psi$.
The $0^{++}$ candidate is a molecule with a mass of $6265.1_{-0.6}^{+0.4}$ MeV/c$^{2}$ and a width of $163_{-7}^{+8}$ MeV, with primary decay channels to $J/\psi J/\psi$ (${\cal B}=65\pm2\%$) and $\eta_c\eta_c$ (${\cal B}=35\pm 2\%$).
Finally, the $2^{++}$ candidate is a resonance that decays entirely to $J/\psi J/\psi$, with mass $6273 \pm 3$ MeV/c$^2$ and with $234_{-13}^{+15}$  MeV. 
It is likely that the experimental signal is a mixture of the two candidates. In order to resolve the different $J^{PC}$ states, we suggest exploring the $\eta_c\eta_c$  channel, which is only accessible for the $0^{++}$ state.

\subsection{$T_{\psi\psi}(6600)$ and $T_{\psi\psi}(6700)$}

The $T_{\psi\psi}(6600)$ tetraquark has been detected in the $J/\psi J/\psi$ invariant mass spectrum at ATLAS~\cite{ATLAS:2022hhx} and CMS~\cite{CMS:2023owd} in proton-proton collision data at $\sqrt{s}=13$ TeV. Its mass and width have been measured to be $6620\pm30$ MeV/c$^2$ and $310\pm90$ MeV, respectively, at ATLAS; and $6552\pm10\pm12$ MeV/c$^2$ and $124^{+32}_{-26}\pm33$ MeV at CMS in a no-interference model and $6638^{+43+16}_{-38-31}$ MeV/c$^2$ and $440^{+230+110}_{-200-240}$ MeV in an interference model. The masses and widths are compatible in the interference model, but the width is significantly smaller in CMS if the no-interference model is used.
In addition, there is a dip in the measured $J/\psi J/\psi$ mass spectrum around $6.75$ GeV, which is not properly accounted for in LHCb's Model I.
To analyse it further, LHCb and CMS used LHCb's Model II, which takes advantage of destructive interference between components and managed to improve the description of the data when a
Breit-Wigner resonance around $6.7$ GeV was added. Although the existence of this state, called $T_{\psi\psi}(6700)$, remains to be confirmed, LHCb determined its mass and width to be $6741\pm6$ MeV/c$^2$ and $288\pm16$ MeV~\cite{LHCb:2020bwg}, respectively, while CMS gave a mass of $6736\pm38$ MeV/c$^2$ and a width of $439\pm 65$ MeV~\cite{CMS:2023owd}.

On the theoretical side, many studies have proposed candidates for the $T_{\psi\psi}(6600)$ and $T_{\psi\psi}(6700)$ tetraquarks, with different properties.
For example, Refs.~\cite{Wang:2022xja} assigned the $T_{\psi\psi}(6600)$ as the first radial excitation of the $0^{++}$ or $1^{+-}$ tetraquark state,
Ref.~\cite{Faustov:2022mvs} identified it as a $0^{++}$ or $2^{++}$ state and, similarly, other studies have candidates with $J^{PC}=0^{++}$, $1^{+-}$ or $2^{++}$~\cite{
Chen:2016jxd,Bedolla:2019zwg,Jin:2020jfc}

Our results show two candidates around $6.6-6.8$ GeV with masses and widths compatible with both the $T_{\psi\psi}(6600)$ and $T_{\psi\psi}(6700)$ and sizable branching ratios to the detection channel $J/\psi J/\psi$.
For example, in $J^{PC}=0^{++}$ we find a resonance with a mass of $6679 \pm 3$ MeV/c$^2$ and a width of $118_{-13}^{+14}$ MeV. Although its mass is slightly larger than the CMS or ATLAS values for the $T_{\psi\psi}(6600)$, its width is compatible with the CMS measurement ($124\pm29\pm34$ MeV). Finally, in the $2^{++}$ sector we have a state with a mass of $6793_{-2}^{+1}$ MeV/c$^2$ and a width of $116_{-10}^{+11}$ MeV, which falls in the energy region of the $T_{\psi\psi}(6700)$, although it is narrower than the actual fits for this state. 
Of course, we need more experimental information to clarify the existence and nature of these states before drawing any conclusions.
Good channels to distinguish these states are the $\eta_c\eta_c$ or $\eta_c\eta_c^\prime$ channels, which are only accessible for the $0^{++}$ state, and the $J/\psi\psi^\prime$ channel, which is only allowed for the $2^{++}$ state.

\subsection{$T_{\psi\psi}(6900)$}

The $T_{\psi\psi}(6900)$ was the first $c\bar cc\bar c$ candidate discovered. It is a narrow structure observed by LHCb in 2020 in the di-$J/\psi$ invariant mass spectrum~\cite{LHCb:2020bwg}.
Its Breit-Wigner mass and width have been determined to be $6905\pm11\pm7$ MeV/c$^2$ and $80\pm 19\pm33$ MeV, respectively, in a fitting scenario without interference, and   
$6886\pm11\pm11$ MeV/c$^2$ and $168\pm33\pm69$ MeV, in a fitting scenario where interference is allowed.
Recently, this structure has been confirmed by CMS~\cite{CMS:2023owd} ($M=6927\pm 9\pm 4$ MeV/c$^2$, $\Gamma=122^{+24}_{-21}\pm18$ MeV) and ATLAS~\cite{ATLAS:2022hhx} ($M=6.87\pm 0.03_{-0.01}^{+0.06}$ GeV/c$^2$ and $\Gamma=0.12\pm 0.04_{-0.01}^{+0.03} $ GeV) in the $J/\psi J/\psi$ mass spectrum. In addition, ATLAS has detected the $T_{\psi\psi}(6900)$ structure in the $J/\psi \psi^\prime$, with BW parameters $6780\pm360$ MeV/c$^2$ and  $390\pm110$  MeV, providing an additional decay channel.

This tetraquark is undoubtedly the most studied. For example, Ref.~\cite{Agaev:2023ruu} assigned it a $0^{++}$ $\chi_{c0}\chi_{c0}$ molecular structure, Ref.~\cite{Wang:2022xja} identified it as a $0^{++}$ second radially-excited tetraquark state, and Ref.~\cite{Karliner:2020dta} concluded that it is most likely a $0^{++}$ radial excitation of a diquark-antidiquark state. 
Other studies agree with the $0^{++}$ assignment~\cite{Lu:2020cns,Chen:2016jxd,Wang:2019rdo,Bedolla:2019zwg,Albuquerque:2020hio}, but leave the door open to other alternatives such as $0^{-+}$, $1^{--}$, $1^{-+}$, $1^{+-}$ or $2^{++}$.

Among all of our candidates in Table~\ref{tab:Tjj2} we can highlight the structures in $0^{++}$, $1^{-+}$ and $2^{++}$ as possible candidates for the $T_{\psi\psi}(6900)$, which are in the $6.8-6.9$ GeV energy region.
We predict two almost degenerate resonances with $J^{PC}=1^{-+}$, whose masses are around $6.9$ GeV/c$^2$ and their widths are $299^{+16}_{-14}$ and $421_{-15}^{+17}$ MeV. These are $J/\psi \psi^\prime$ states in a relative $P$-wave, one of them mixed with the $\eta_c\eta_c^\prime$ channel, thus they are candidates to the ATLAS sign of the $T_{\psi\psi}(6900)$ states. Unlike the $0^{++}$ and $2^{++}$ candidates, one of the former $1^{-+}$ resonances can also decay to the $\eta_c \eta_c^\prime$ channel, so this is a
good channel to evaluate its existence. 

In the $J^{PC}=0^{++}$ sector we also have a signal in the $J/\psi J/\psi$ and $J/\psi\psi^\prime$ mass spectrum, due to a virtual state below the $J/\psi\psi^\prime$ threshold, in the second Riemann sheet. Its mass is $6782_{-3}^{+2}$ MeV/c$^2$ and its width $18_{-6}^{+9}$ MeV, although as it is a virtual state its width cannot be directly compared with the Breit-Wigner properties experimentally measured. It mainly decay to $J/\psi J/\psi$ (${\cal B}=70_{-11}^{+1}\%$), and also to $\eta_c\eta_c^\prime$ (${\cal B}=21_{-2}^{+1}\%$), which could be a good detection channel.

Finally, the $2^{++}$ candidate is a resonance with a mass of $6793_{-2}^{+1}$ MeV/c$^2$ and a width of $116_{-10}^{+11}$ MeV. It is a state which mainly decays to $J/\psi J/\psi$ (${\cal B}=68\pm 1\%$) and $J/\psi \psi^\prime$ (${\cal B}=32\pm 1\%$). Its width is compatible with the experimental data from LHCb, CMS and ATLAS in di-$J/\psi$ channel, whereas its mass is slightly smaller.

\subsection{$T_{\psi\psi}(7200)$}

In addition to the above $T_{\psi\psi}(6900)$ state, the LHCb Collaboration suggested a broad structure peaking at about $7.2$ GeV, later named $T_{\psi\psi}(7200)$. 
In 2022, the CMS~\cite{CMS:2023owd} and ATLAS~\cite{ATLAS:2022hhx} collaborations provided its Breit-Wigner properties, measured from the $J/\psi J/\psi$ and $J/\psi \psi^\prime$ mass spectra data, respectively. Its mass was determined to be $7287^{+20}_{-18}\pm5$ MeV/c$^2$ (CMS) and $7220\pm30$ MeV/c$^2$ (ATLAS), while its width was measured to be $95^{+59}_{-40}\pm19$ MeV (CMS) and $100^{+130}_{-70}$ MeV (ATLAS).
From a theoretical point of view, this state was mostly identified as a $0^{++}$ structure~\cite{Karliner:2020dta,Wang:2022xja,Faustov:2022mvs,Bedolla:2019zwg,Jin:2020jfc}, but other alternatives such as $1^+$ or $2^+$ were suggested~\cite{Faustov:2022mvs,Wang:2019rdo,Bedolla:2019zwg,Jin:2020jfc}. 

For this state, we predict one near virtual candidate with $0^{++}$ quantum numbers, around $7.3$ GeV/c$^2$.
This $0^{++}$ state has a mass of $7276_{-1}^{+2}$ MeV/c$^2$ and a width of $35_{-13}^{+11}$. It mainly decays to $J/\psi J/\psi$, $\eta_c\eta_c$ and $\eta_c\eta_c^\prime$, with a small branching to $J/\psi\psi^\prime$.
There are other states close in mass, such as a $1^{+-}$ virtual state that has a mass of $7303_{-1}^{+3}$ MeV/c$^2$ and a width of $25_{-14}^{+11}$ MeV, but this state only decays to $\eta_c J/\psi$, $\eta_c \psi^\prime$ and $\eta_c^\prime J/\psi$, so it cannot be the $T_{\psi\psi}(7200)$. In $0^{++}$ and $2^{++}$ there are also two slightly heavier virtual states, with masses and widths ($7349.6_{-1.0}^{+0.3}$ MeV/c$^2$, $67_{-9}^{+7}$ MeV) and ($7342 \pm 2$ MeV/c$^2$,$94.9_{-0.9}^{+1.0}$ MeV) respectively, which can decay to $J/\psi J/\psi$ and $J/\psi \psi^\prime$. We cannot discard that the experimental $T_{\psi\psi}(7200)$ is a mixture of the above $0^{++}$ and $2^{++}$ virtual states.
We want to remark here that the position and width of the virtual poles cannot be directly compared to the Breit-Wigner parameters as measured by the LHCb, CMS and ATLAS collaborations, as the virtuals are in an unphysical sheet and we only see them as bumps above the nearest thresholds.

\begin{turnpage}
\begin{table}[!t]
\caption{\label{tab:Tjj2} Coupled-channels calculation of the $J^P=0^\pm$, $1^\pm$ and $2^\pm$ $cc \bar c \bar c$ sectors ($T_{\psi\psi}$ states) as meson-meson molecules, including the channels detailed in Table~\ref{tab:thres}. Errors are estimated by varying the strength of the potential by $\pm10\%$. \emph{$1^{st}$ column:} Pole's quantum numbers; \emph{$2^{nd}$ column:} Pole's mass in MeV/c$^2$; \emph{$3^{rd}$ column:} Pole's width in MeV; \emph{$4^{th}$-$13^{th}$ columns:}  Branching ratios in \%. States with a dagger before their mass are virtual states, defined as poles in the second Riemann sheet below their closest threshold.}
\begin{ruledtabular}
\begin{tabular}{lllllllllllll}
 $J^{PC}$ & $M_{\text{pole}}$ & $\Gamma_{\text{pole}}$  & ${\cal B}_{\eta_c\eta_c}$ & ${\cal B}_{\eta_cJ/{\psi}}$ & ${\cal B}_{J/{\psi}J/{\psi}}$ & ${\cal B}_{\eta_c\eta_c^{\prime}}$ & ${\cal B}_{\eta_c\psi^{\prime}}$ & ${\cal B}_{\eta_c^{\prime}J/{\psi}}$ & ${\cal B}_{J/{\psi}\psi^{\prime}}$ & ${\cal B}_{\eta_c^{\prime}\eta_c^{\prime}}$ & ${\cal B}_{\eta_c^{\prime}\psi^{\prime}}$ & ${\cal B}_{\psi^{\prime}\psi^{\prime}}$ \\[1ex]
\hline
 $ 0^{--}$ & $6741_{-7}^{+6}$ & $546_{-13}^{+14}$  & $0$  & $0$  & $0$  & $0$  & $100$  & $0$  & $0$  & $0$  & $0$  & $0$  \\
        & $6821_{-4}^{+3}$ & $407_{-16}^{+18}$  & $0$  & $0$  & $0$  & $0$  & $50.7 \pm 0.3$  & $48.9_{-0.3}^{+0.4}$  & $0$  & $0$  & $0$  & $0$  \\[1ex]
$ 0^{++}$ & $6088_{-4}^{+3}$ & $245_{-18}^{+20}$  & $100$  & $0$  & $0$  & $0$  & $0$  & $0$  & $0$  & $0$  & $0$  & $0$  \\
       & $6265.1_{-0.6}^{+0.4}$ & $163_{-7}^{+8}$  & $35 \pm 2$  & $0$  & $65 \pm 2$  & $0$  & $0$  & $0$  & $0$  & $0$  & $0$  & $0$  \\
       & $6679 \pm 3$ & $118_{-13}^{+14}$  & $28.3_{-0.8}^{+0.3}$  & $0$  & $8_{-1}^{+2}$  & $64 \pm 2$  & $0$  & $0$  & $0$  & $0$  & $0$  & $0$  \\
       & $7047_{-7}^{+9}$ & $375_{-20}^{+24}$  & $14 \pm 0$  & $0$  & $56.1_{-0.1}^{+0.5}$  & $7 \pm 0$  & $0$  & $0$  & $22.8_{-0.5}^{+0.8}$  & $0$  & $0$  & $0$  \\
       & $7503_{-2}^{+3}$ & $227_{-5}^{+9}$  & $0_{-0}^{+2}$  & $0$  & $18_{-8}^{+9}$  & $19_{-3}^{+2}$  & $0$  & $0$  & $16 \pm 3$  & $0.9 \pm 0.5$  & $0$  & $46_{-3}^{+1}$  \\
       & $7587 \pm 4$ & $449 \pm 1$  & $7.5_{-0.0}^{+0.2}$  & $0$  & $20_{-4}^{+5}$  & $2 \pm 1$  & $0$  & $0$  & $1.9 \pm 0.2$  & $64_{-5}^{+4}$  & $0$  & $4.0_{-1.0}^{+0.9}$  \\
       & $^\dagger$$6782_{-3}^{+2}$  & $18_{-6}^{+9}$  & $8_{-2}^{+0}$  & 0  & $70_{-11}^{+1}$  & $21_{-2}^{+1}$  & 0  & 0  & $0_{-0}^{+15}$  & 0  & 0  & 0  \\
       & $^\dagger$$7276_{-1}^{+2}$ & $35_{-13}^{+11}$  & $24_{-3}^{+2}$  & 0  & $37 \pm 2$  & $34_{-4}^{+5}$  & 0  & 0  & $5 \pm 0$  & 0  & 0  & 0  \\
       & $^\dagger$$7349.6_{-1.0}^{+0.3}$  & $67_{-9}^{+7}$  & $17.3_{-1.0}^{+0.1}$  & 0  & $45_{-13}^{+14}$  & $7_{-1}^{+2}$  & 0  & 0  & $11_{-8}^{+7}$  & $20_{-4}^{+5}$  & 0  & 0  \\[1ex]
 $ 1^{-+}$ & $6747 \pm 5$ & $450_{-18}^{+19}$  & $0$  & $0$  & $0$  & $100$  & $0$  & $0$  & $0$  & $0$  & $0$  & $0$  \\
       & $6888 \pm 1$ & $299_{-14}^{+16}$  & $0$  & $0$  & $0$  & $24 \pm 1$  & $0$  & $0$  & $76 \pm 1$  & $0$  & $0$  & $0$  \\
       & $6884_{-5}^{+4}$ & $421_{-15}^{+17}$  & $0$  & $0$  & $0$  & $0$  & $0$  & $0$  & $100$  & $0$  & $0$  & $0$  \\
[1ex]
 $ 1^{--}$  & $6743_{-7}^{+6}$ & $545_{-13}^{+14}$  & $0$  & $0$  & $0$  & $0$  & $100$  & $0$  & $0$  & $0$  & $0$  & $0$  \\
        & $6822_{-4}^{+3}$ & $405_{-16}^{+18}$  & $0$  & $0$  & $0$  & $0$  & $50.7 \pm 0.3$  & $48.9_{-0.4}^{+0.3}$  & $0$  & $0$  & $0$  & $0$  \\
[1ex]
 $ 1^{+-}$ & $6173 \pm 2$ & $212_{-14}^{+16}$  & $0$  & $100$  & $0$  & $0$  & $0$  & $0$  & $0$  & $0$  & $0$  & $0$  \\
        & $6741.4_{-0.7}^{+0.3}$ & $116_{-8}^{+9}$  & $0$  & $41 \pm 1$  & $0$  & $0$  & $42.0_{-0.6}^{+0.8}$  & $17 \pm 2$  & $0$  & $0$  & $0$  & $0$  \\
        & $7528.6_{-0.2}^{+0.0}$ & $288_{-10}^{+12}$  & $0$  & $26_{-4}^{+3}$  & $0$  & $0$  & $1.0_{-0.3}^{+0.5}$  & $14 \pm 4$  & $0$  & $0$  & $59.1 \pm 0.9$  & $0$  \\
        & $^\dagger$$6659_{-1}^{+0}$ & $182 \pm 14$  & $0$  & $100$  & $0$  & $0$  & $0$  & $0$  & $0$  & $0$  & $0$  & $0$  \\
        & $^\dagger$$7303_{-1}^{+3}$ & $25_{-14}^{+11}$  & $0$  & $15 \pm 1$  & $0$  & $0$  & $46_{-2}^{+1}$  & $38_{-0}^{+5}$  & $0$  & $0$  & $0$  & $0$  \\[1ex]
 $ 2^{--}$ & $6742_{-7}^{+6}$ & $545_{-13}^{+14}$  & $0$  & $0$  & $0$  & $0$  & $100$  & $0$  & $0$  & $0$  & $0$  & $0$  \\
        & $6822_{-4}^{+3}$ & $406_{-16}^{+18}$  & $0$  & $0$  & $0$  & $0$  & $50.7 \pm 0.3$  & $48.9_{-0.3}^{+0.4}$  & $0$  & $0$  & $0$  & $0$  \\[1ex]
$ 2^{++}$ & $6273 \pm 3$ & $234_{-13}^{+15}$  & $0$  & $0$  & $100$  & $0$  & $0$  & $0$  & $0$  & $0$  & $0$  & $0$  \\
& $6793_{-2}^{+1}$ & $116_{-10}^{+11}$  & $0$  & $0$  & $68_{-1}^{+2}$  & $0$  & $0$  & $0$  & $32_{-2}^{+1}$  & $0$  & $0$  & $0$  \\
& $7143 \pm 6$ & $524_{-18}^{+19}$  & $0$  & $0$  & $56 \pm 2$  & $0$  & $0$  & $0$  & $44 \pm 2$  & $0$  & $0$  & $0$  \\       
       & $7515 \pm 3$ & $321 \pm 14$  & $0$  & $0$  & $59_{-1}^{+0}$  & $0$  & $0$  & $0$  & $5.6_{-0.3}^{+0.6}$  & $0$  & $0$  & $35_{-0}^{+1}$  \\
 & $^\dagger$$7342 \pm 2$ & $94.9_{-0.9}^{+1.0}$  & $0$  & $0$  & $71.5_{-0.1}^{+0.4}$  & $0$  & $0$  & $0$  & $28.5_{-0.4}^{+0.1}$  & $0$  & $0$  & $0$  \\
       [1ex]
$2^{+-}$    & $7616_{-7}^{+6}$ & $531_{-11}^{+12}$  & $0$  & $0$  & $0$  & $0$  & $1.3 \pm 0.2$  & $2.0 \pm 0.2$  & $0$  & $0$  & $96.6 \pm 0.4$  & $0$  \\
\end{tabular}
\end{ruledtabular}
\end{table}
\end{turnpage}


\begin{table*}[b!]
 \caption{\label{tab:summary} Summary of the tentative assignment of our theoretical  poles with respect to the experimental $T_{\psi\psi}$  states. The $T_{\psi\psi}(6700)$ has been suggested by CMS, but it is still not confirmed. See text and experimental references for more details. \emph{$1^{st}$ column}: Assigned name, \emph{$2^{nd}$ column}: Collaboration and reference of the experimental study. \emph{$3^{rd}$ column}: Detection channel, \emph{$4^{th}$ column}: Experimental mass in MeV/c$^2$ (first error is statistical, second error is systematic). \emph{$5^{th}$ column}: Experimental width in MeV. \emph{$6^{th}$ column}: Quantum numbers of the theoretical candidate, \emph{$7^{th}$ column}: Theoretical mass of the candidate (in MeV/c$^2$), \emph{$8^{th}$ column}: Theoretical width of the candidate (in MeV). States with a dagger before their mass are virtual states, defined as poles in the second Riemann sheet below their closest threshold. }
 \begin{ruledtabular}
\begin{tabular}{llllllll}
 State & Coll. & Channel& $M_{\rm exp}$ & $\Gamma_{\rm exp}$ & $J^{PC}$ & $M_{\rm pole}$ & $\Gamma_{\rm pole}$ \\ \hline
 $T_{\psi\psi}(6200)$  & ATLAS~\cite{ATLAS:2022hhx} & di-$J/\psi$& $6220\pm50$ & $310\pm120$ & $0^{++}$ & $6265.1_{-0.6}^{+0.4}$ & $163_{-7}^{+8}$  \\
                    &   &       &             &             & $ 2^{++}$ & $6273 \pm 3$ & $234_{-13}^{+15}$  \\ [1ex]
 $T_{\psi\psi}(6600)$  & ATLAS~\cite{ATLAS:2022hhx} & di-$J/\psi$& $6620\pm30$ & $310\pm90$  & $0^{++}$ & $6679 \pm 3$ & $118_{-13}^{+14}$  \\
                       & CMS (Model I)~\cite{CMS:2023owd} & di-$J/\psi$& $6552\pm10\pm12$ & $124^{+32}_{-26}\pm33$  &  &  &  \\
                       & CMS (Model II)~\cite{CMS:2023owd} & di-$J/\psi$& $6638^{+43+16}_{-38-31}$ & $440^{+230+110}_{-200-240}$  &  &  &  \\[1ex]
 $T_{\psi\psi}(6700)?$ & LHCb~\cite{LHCb:2020bwg} & di-$J/\psi$& $6741\pm6$ & $288\pm 16$  & $0^{++}$ & $6679 \pm 3$ & $118_{-13}^{+14}$  \\
 & CMS~\cite{CMS:2023owd}  & di-$J/\psi$&  $6736\pm38$ & $439\pm 65$  & $2^{++}$ & $6793_{-2}^{+1}$ & $116_{-10}^{+11}$  \\  [1ex]
 $T_{\psi\psi}(6900)$ & LHCb (Model I)~\cite{LHCb:2020bwg} & di-$J/\psi$& $6905\pm11\pm7$ & $80\pm 19\pm33$   & $0^{++}$ & $^\dagger$$6782_{-3}^{+2}$  & $18_{-6}^{+9}$ \\ 
                      & LHCb (Model II)~\cite{LHCb:2020bwg} & di-$J/\psi$& $6886\pm11\pm11$ & $168\pm33\pm69$  & $1^{-+}$ & $6888\pm1$  & $299_{-14}^{+16}$ \\ 
                      & CMS (Model I)~\cite{CMS:2023owd} & di-$J/\psi$& $6927\pm9\pm4$ & $122^{+24}_{-21}\pm18$  & $1^{-+}$ & $6884_{-5}^{+4}$ & $421_{-15}^{+17}$ \\ 
                      & CMS (Model II)~\cite{CMS:2023owd} & di-$J/\psi$& $6847^{+44+48}_{-28-20}$ & $191^{+66+25}_{-49-17}$  & $2^{++}$ & $6793_{-2}^{+1}$ & $116_{-10}^{+11}$\\ 
                      & ATLAS~\cite{ATLAS:2022hhx} & di-$J/\psi$& $6870\pm30$ & $120\pm40$  &  &  & \\
                      & ATLAS~\cite{ATLAS:2022hhx}& $J/\psi \psi^\prime$& $6780\pm360$ &  $390\pm110$  &  &  &  \\ [1ex]
 $T_{\psi\psi}(7200)$ & CMS (Model I)~\cite{CMS:2023owd} & di-$J/\psi$& $7287^{+20}_{-18}\pm5$ & $95^{+59}_{-40}\pm19$  & $0^{++}$ & $^\dagger$$7276_{-1}^{+2}$  & $35_{-13}^{+11}$ \\
 & CMS (Model II)~\cite{CMS:2023owd} & di-$J/\psi$& $7134^{+48+41}_{-25-15}$ & $97^{+40+29}_{-29-26}$  & $0^{++}$  & $^\dagger$$7349.6_{-1.0}^{+0.3}$  & $67_{-9}^{+7}$\\
                      & ATLAS~\cite{ATLAS:2022hhx} & $J/\psi \psi^\prime$& $7220\pm30$ & $100^{+130}_{-70}$ & $2^{++}$  & $^\dagger$$7342 \pm 2$ & $94.9_{-0.9}^{+1.0}$ \\
\end{tabular}
\end{ruledtabular}
\end{table*}

\section{SUMMARY}
\label{sec:summary}

\begin{figure}[htb!]
\includegraphics[width=.55\textwidth]{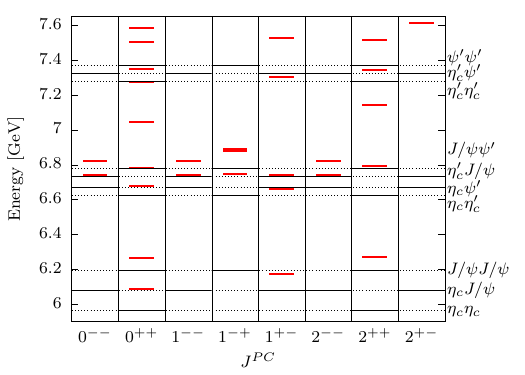}
\caption{\label{fig1} Summary of the $T_{\psi\psi}$ candidates found in this work (red lines). The opened (closed) $c\bar c-c\bar c$ thresholds are shown as horizontal solid (dashed) lines. See Table~\ref{tab:Tjj2} for more details. }
\end{figure}

In this study we have analysed the $c\bar c- c\bar c$ system in a coupled-channels calculation of the $J^P=0^\pm$, $1^\pm$ and $2^\pm$ sectors, including the channels $\eta_c\eta_c$, $\eta_c J/\psi$, $J/\psi J/\psi$, $\eta_c\eta_c^\prime$, $\eta_c\psi^\prime$, $\eta_c^\prime J/\psi$, $J/\psi \psi^\prime$, $\eta_c^\prime\eta_c^\prime$, $\eta_c^\prime\psi^\prime$ and $\psi^\prime\psi^\prime$ (that's it, all channels containing a $J/\psi$, $\psi^\prime$, $\eta_c$ and $\eta_c^\prime$), with the partial waves detailed in table~\ref{tab:thres}.
We have searched for poles in the scattering matrix and found $29$ states with masses between $6.1$ and $7.6$ GeV/c$^2$ in different $J^{P}$ sectors (see Fig.~\ref{fig1}). In particular, we find $2$ states in $0^-$, $9$ in $0^+$, $5$ in $1^-$, $5$ in $1^+$, $2$ in $2^-$ and $6$ in $2^+$. Their masses, widths and branching ratios have been studied (see table~\ref{tab:Tjj2}), finding candidates for the experimental $T_{\psi\psi}(6200)$, $T_{\psi\psi}(6600)$, $T_{\psi\psi}(6700)$, $T_{\psi\psi}(6900)$ and $T_{\psi\psi}(7200)$ tetraquarks.

A summary of our tentative assignments compared to the current experimental $cc\bar c\bar c$ candidates is given in Table~\ref{tab:summary}. We have discussed different detection channels that could help to discriminate between different candidates, and analysed the best strategies to search for the rest of the predicted $T_{\psi\psi}$ states.

\begin{acknowledgments}
This work has been partially funded by
EU Horizon 2020 research and innovation program, STRONG-2020 project, under grant agreement no. 824093 and
Ministerio Espa\~nol de Ciencia e Innovaci\'on, grant no. PID2019-105439GB-C22.
\end{acknowledgments}


\bibliography{draftJpsiJpsi}

\end{document}